\documentstyle[12pt,bezier]{article}
\input psfig

\newcommand{\bra}[1]{\langle #1 |}
\newcommand{\ket}[1]{| #1 \rangle}

\newcommand{\braket}[2]{\langle #1 | #2 \rangle}
\newcommand{\eq}[1]{eq.(\ref{#1})}
\newcommand{\dpar}[2]{\frac{\partial #1}{\partial #2}}

\def\e{\mbox{e}}

\begin{document}
\title{Baby universes and energy (non-)conservation in (1+1)-dimensional
dilaton gravity}
\author{
   V.~A.~Rubakov\\
    {\small \em Institute for Nuclear Research of the Russian Academy of
  Sciences,}\\
  {\small \em 60th October Anniversary prospect 7a, Moscow 117312}\\
  }
%\date{May 1995}
\maketitle

hep-th/9611171

\begin{abstract}
We consider branching of baby universes off parent one in (1+1)-dimensional
dilaton gravity with 24 types of conformal matter fields. This theory is
equivalent to string theory in a certain background in $D=26$-dimensional
target space, so this process may be also viewed as the emission of a light
string state by a heavy string. We find that bare energy is not conserved in
(1+1) dimensions due to the emission of baby universes, and that the 
probability of this process is finite even for local distribution of matter in
the parent universe. We present a scenario suggesting that the 
non-conservation of bare energy may be consistent with the locality of the
baby universe emission process in (1+1) dimensions {\it and} the existence of
the long ranged dilaton field whose source is bare energy. This scenario 
involves the generation of longitudinal gravitational waves in the
parent universe.
\end{abstract}

%\vskip .5 in

%hep-th/9611171

\newpage
\section{Introduction}
Generally covariant (1+1)-dimensional theories provide convenient 
framework for considering various suspected properties of quantum gravity
in (3+1) dimensions (for reviews see, e.g., 
refs.\cite{Ginsparg,StromingerReview}).
In particular, the long standing issue of the possible role of topology
changing transitions and baby 
universes\cite{Hawking1,LRT,GS1,Coleman1,GS2,Banks} may be naturally 
discussed in this
framework. A special feature of (1+1) dimensions which is useful for the
study of baby universes/wormholes is that some models admit their 
interpretation as string theories in higher dimensional target space.

The simplest model of this sort is literally the theory of closed 
strings in the Minkowski target space of critical 
dimensions\cite{Hawking2,LyHa,VR,Nirov}. Indeed,
macroscopic and microscopic string states may be interpreted as 
(1+1)-dimensional parent and baby universes, respectively. One feature 
inherent in that model is that the emission of a baby universe always requires
non-conservation of bare energy in the parent universe\footnote{It has been
argued \cite{VR} that the emission of  baby strings should lead to the
loss of quantum coherence for one-dimensional observer at the parent string.
Independently, it has been argued on general grounds\cite{BanksPS}
that the energy non-conservation is inevitable in modifications of
quantum mechanics allowing for the loss of quantum coherence (see,
however, refs. \cite{EllisEnergy,Unruh}). So, energy non-conservation in the
stringy model of ref. \cite{VR} may not be 
too surprising.}\cite{VR,VRFilippov}. This non-conservation does not
lead to any drastic consequences in the simplest stringy model; in particular,
the rate of emission of baby strings is finite in the limit of infinite
size of the parent string\cite{Nirov}.

However, one important feature present in (3+1)-dimensional gravity 
is missing in the simplest stringy model. Namely, in (3+1) dimensions there
exists long ranged gravitational field whose source in energy--momentum
(Newton's gravity law), while there is no such field in that stringy model.
Intuitively, one may suspect that the existence of the long ranged 
field associated with energy and momentum may be an obstacle to energy 
non-conservation in local processes like the emission of baby universes.
To address this issue, more refined model than that of closed strings
in critical dimensions, is needed.

A particularly simple (1+1)-dimensional model where the mass (energy--momentum
of matter fields) produces long range effects, is the dilaton gravity 
with matter
that has been widely discussed from the point of view of black hole
physics\cite{CGHS} (for careful analysis of the notion of
ADM mass in that model see refs.\cite{BilalKogan,Bilal}). Here we take 
a different attitude and consider the emission of baby universes, so we
simplify the model as much as possible. In particular, we set the number of
matter fields equal to 24 and the (1+1)-dimensional cosmological
constant to zero. As stressed in refs.\cite{Tseytlin1,Tseytlin2},
this model is equivalent to bosonic string theory (in critical dimension 
$D=26$ of target space) in linear Dilaton background (to distinguish
between dilaton fields in (1+1)-dimensional world and in $D$-dimensional
target space, we call the former ``dilaton'' and the latter ``Dilaton'',
respectively). Hence, the emission of a (1+1)-dimensional baby universe
by a parent universe
has an interpretation from the $D$-dimensional point of view as the emission 
of a light string state by a highly excited string state, in complete 
analogy to ref.\cite{VR}. This process, in the leading order of
string perturbation theory, is tractable both qualitatively and 
quantitatively; in particular, one is able to analyze whether it is
accompanied by non-conservation of (bare) energy in (1+1) dimensions and 
whether its rate  is large (unsuppressed) when baby universes are 
emitted locally in (1+1)-dimensional parent universe. The discussion of 
these points is the main purpose of this paper.

The outline of the paper is as follows. In section 2 we describe the model
and some of its classical solutions in (1+1) dimensions. For technical
reasons, the quantum version is conventiently constructed for the case
of closed (1+1)-dimensional universe, so we present in section 2 some
classical solutions in the closed universe. This discussion will be useful to
understand that the (bare) energy of localized distributions of matter
fields still produces long ranged effects, at least in some gauges, even 
though the total energy of the closed universe is always zero. In section 3 we
outline the quantum version of this model, which is known for some time
(see, e.g., refs.\cite{Verlinde1,Verlinde2,Kazama1,Kazama2}), construct 
the states of parent universes (DDF states) and vertex operators 
corresponding to the emission of baby universes. Section 4 contains the main
results of this paper. We consider the simplest DDF state of the parent 
universe, which can be interpreted as containing just two dressed matter 
``particles'', and analyze the emission of baby universes by this state in
the lowest order of string perturbation theory. We find that this 
emission always occurs with non-conservation of energy of matter in 
(1+1)-dimensional parent universe. We then proceed to  the explicit
calculation of  the emission rate.  Surprisingly, we find that the
rate is  unsuppressed even for localized distribution of matter in
the parent universe, in spite of the long range field this matter produces.
In section 5 we conclude by presenting a scenario showing that the
non-conservation of bare energy of matter may be consistent both with 
locality of the emission process and with the presence of long ranged field;
in this scenario, the ADM mass is conserved at the expense of the
generation of ``longitudinal'' gravitational waves due to the emission of
a baby universe.

\section{Model and classical solutions}
\subsection{The model}

The action for the simplest version of (1+1)-dimensional
dilaton gravity with conformal matter can be written in the form
similar to ref.\cite{Russo},
\begin{equation}
  S= -\frac{1}{\pi} \int~d^2\sigma~ \sqrt{-g}\left(-\frac{\gamma^2}{4}\phi R
          +g^{\alpha\beta}\partial_{\alpha}f^i \partial_{\beta}f^i \right)
\label{8+}
\end{equation}
where $\phi$ is the dilaton field, $f^i$ are matter fields, $i=1,\dots,24$,
and $\gamma$ is a positive coupling constant analogous to the Planck mass
of (3+1)-dimensional gravity. The coupling constant $\gamma$ may be absorbed
into the dilaton field, but we will not do this for book-keeping purposes.
Both in infinite space and in the closed
(1+1)-dimensional universe the field equations are simplified in the
conformal gauge
\[
       g_{\alpha\beta} = \e^{2\rho}\eta_{\alpha\beta}
\]
where $\eta$ is the Minkowskian metrics in (1+1) dimensions. In this gauge,
the fields $\rho$, $\phi$ and $f^i$ obey massless free field equations.
There are also constraints
\begin{equation}
   -\frac{1}{2}\gamma^2\left(\partial_{\pm}\phi \partial_{\pm}\rho
         - \frac{1}{2}\partial_{\pm}^2 \phi\right) +
    \frac{1}{2} \left(\partial_{\pm}{\bf f}\right)^2 = 0
\label{8*}
\end{equation}
ensuring that the total energy--momentum tensor vanishes.

\subsection{Solutions in infinite space}

Let us outline some classical solutions in this model. We begin with 
the case of infinite one-dimensional space, $\sigma^1 \in (-\infty,+\infty)$,
and consider localized distributions of matter. In this case one can 
further specify the gauge and choose
\begin{equation}
  \rho=0
\label{9b}
\end{equation}
so that the space-time is flat. Equation (\ref{8*}) then determines
 the dilaton field $\phi$ for a given matter distribution. Indeed, 
the solution to \eq{8*} is, up to an arbitrary linear function of coordinates,
\[
   \phi(\sigma) = \phi_{+}(\sigma_{+}) +
                   \phi_{-}(\sigma_{-}) 
\]
with
\begin{equation}
   \phi_{\pm} = -\frac{1}{\gamma^2} \int~d\sigma_{\pm}'~
                |\sigma_{\pm} - \sigma'_{\pm}|~
                (\partial_{\pm}{\bf f})^2 (\sigma'_{\pm})
\label{9a}
\end{equation}
Hence, the energy--momentum of matter fields produces long ranged dilaton 
field which has linear  behavior at large $|\sigma^1|$. In
particular, the ADM mass can be defined as follows,
\begin{equation}
      \mu_{ADM} = - \frac{\gamma^2}{2\pi}
              \left[\dpar{\phi}{\sigma^1}(\sigma^1 \to +\infty) -
               \dpar{\phi}{\sigma^1}(\sigma^1 \to -\infty) \right]
\label{9c}
\end{equation}
In virtue of \eq{9a} it is equal to
\[
     \mu_{ADM} = \int_{-\infty}^{+\infty}~d\sigma^1~\varepsilon_M(\sigma)
\]
where
\[
    \varepsilon_M = \frac{1}{2\pi}\left[ (\partial_0 {\bf f})^2
                       + (\partial_1 {\bf f})^2 \right]
\]
is the energy density of matter which we will often call bare energy
density \footnote{Note somewhat 
unconventional factor $1/\pi$ in the matter action
in \eq{8+}.}.

The dilaton field produced by two lumps of matter of equal energy and 
opposite momenta, moving towards each other (or from each other) with the speed
of light, is shown in fig.1. Needless to say, the linear dependence 
of $\phi$ on $\sigma^1$ at large $|\sigma^1|$ is nothing but the Coulomb
behavior of long ranged field in one-dimensional space. In this respect
the dilaton field in (1+1) dimensions is analogous to gravitational field
of Newton's law in (3+1) dimensions.

\subsection{Solutions in closed space}

For technical reasons, the quantum version of this model
is conveniently formulated in closed one-dimensional space. So, it is
instructive to consider classical solutions in the closed universe.
Let us study the classical theory (\ref{8+}) on a circle 
\[
\sigma^1 \in \left(-\frac{\pi}{2}, +\frac{\pi}{2} \right)
\]
The absolute length of the circle is irrelevant as we are dealing 
with scale-invariant action; what will matter is the relative size of matter
distribution to the length of the universe. The gauge (\ref{9b}) is no
longer possible in the closed universe;
the closest analog is the gauge in which the universe  contracts (or expands)
homogeneously,
\begin{equation}
       \rho = - \frac{1}{\gamma} P^{(-)} \sigma^0
\label{9+}
\end{equation}
where $P^{(-)}$ is some constant; our choice of normalization
and notation will become clear later. In this gauge, the constraints 
(\ref{8*}) may again be used to determine the dilaton field for a 
given distribution of matter, provided the total spatial
momentum of matter vanishes,
\begin{equation}
    \int_{-\pi/2}^{+\pi/2}~d\sigma^1~(\partial_{+}{\bf f})^2
        = \int_{-\pi/2}^{+\pi/2}~d\sigma^1~(\partial_{-}{\bf f})^2
\label{10++}
\end{equation}
The constraints (\ref{8*}) in the gauge (\ref{9+}) read
\begin{equation}
    - \frac{1}{4}\gamma P^{(-)} \partial_{\pm}\phi -
     \frac{1}{4} \gamma^2\partial^2_{\pm}\phi =
     \frac{1}{2} (\partial_{\pm}{\bf f})^2
\label{12a}
\end{equation}
Equation (\ref{10++}) is an immediate consequence of these constraints and the
periodicity of the dilaton field in $\sigma^1$.

For a given matter distribution, the solution to \eq{12a} which is 
periodic in $\sigma^1$ with period $\pi$ is
\begin{equation}
    \phi(\sigma^0,\sigma^1) = \phi_{+}(\sigma_{+})+
                              \phi_{-}(\sigma_{-})
\label{10*}
\end{equation}
where
\begin{equation}
   \phi_{\pm}(\sigma_{\pm}) = \int~ d\sigma_{\pm}~
                  G(\sigma_{\pm},\sigma'_{\pm}) ~\frac{1}{2} 
                   (\partial_{\pm}{\bf f})^2
\label{10**}
\end{equation}
and the Green function of \eq{12a} obeys the periodicity condition
\[
    \partial_u G(u+\pi,u') = \partial_u G(u,u')
\]
At
$u \in (-\pi/2, \pi/2)$, $u' \in (-\pi/2, \pi/2)$ we have explicitly
\begin{equation}
         G(u,u') =
      \frac{2}{\gamma P^{(-)} \sinh (\pi P^{(-)}/2\gamma)}
      \exp \left[\frac{\pi P^{(-)}}{2\gamma}\epsilon (u-u')\right]
      \left( \exp \left[-\frac{P^{(-)}}{\gamma}(u-u')\right] - 1 \right)
\label{10+}
\end{equation}
where $\epsilon (u-u')$ is the usual step function.

It is instructive to consider the case $\gamma \gg 1$ (large ``Planck mass'')
and study two narrow pulses of matter moving left and right and colliding at
$\sigma^1 = 0$. These pulses may be
approximated by the delta-function distribution,
\begin{equation}
     (\partial_{\pm} {\bf f})^2 = \frac{\pi}{2} \mu \delta (\sigma_{\pm})
\label{13a}
\end{equation}
where the normalization is such that the constant $\mu$ would coincide with the
ADM mass had the universe infinite size,
\begin{equation}
 \mu =  \int~d\sigma^1~
\frac{1}{2\pi}\left[ (\partial_0 {\bf f})^2
                       + (\partial_1 {\bf f})^2 \right]
\label{13d}
\end{equation}
In this case \eq{10**} has particularly simple form (again at 
$|\sigma_{\pm}|< \pi/2$)
\begin{equation}
    \phi_{\pm} = -\frac{P^{(+)}}{2\gamma} \sigma_{\pm}
                 - \frac{\pi}{2\gamma^2} \mu |\sigma_{\pm}| 
                 + \frac{\mu}{2\gamma^2} \sigma_{\pm}^2 + O(\gamma^{-3})
\label{13b}
\end{equation}
where
\begin{equation}
   P^{(+)} = \frac{2\mu}{P^{(-)}}
\label{14b}
\end{equation}
When the pulses are close to each other, one has for the total dilaton field
at $|\sigma_{\pm}| \ll 1$
\begin{equation}  
 \phi = -\frac{P^{(+)}}{\gamma}\sigma^0
                 - \frac{\pi}{2\gamma^2} \mu (|\sigma_{+}| + |\sigma_{-}|)
                 + \dots
\label{14a}
\end{equation}
The first term in \eq{14a} describes spatially homogeneous component of the
dilaton field, while the second term shows precisely the same 
Coulomb behavior as in the case of infinite space. In fact, the latter term
coincides with the expression (\ref{9a}) (for narrow matter pulses as
defined in \eq{13a}).
The terms omitted in \eq{14a} are of order of $\gamma^{-2}$ and they become 
important at $|\sigma_{\pm}|\sim 1$; in particular, they ensure that 
$\phi(\sigma^0,\sigma^1)$ flattens out and has vanishing spatial derivatives
at $\sigma^1= \pm \pi /2$. The behavior of the dilaton field generated
by narrow pulses of matter, which are close to each other, is
schematically shown in fig.2.

Even though the ADM mass is, strictly speaking, zero in the closed
universe, narrow pulses of matter produce the dilaton field that shows
Coulomb behavior not far away from the pulses. In this sense one can 
still use the  notion of bare energy in the gauge (\ref{9+}). This bare 
energy (the energy of matter) can be observed by a one-dimensional
observer by measuring the dilaton field outside the pulses but at
distances small compared to the size of the universe. With these 
reservations, the formula for the ADM mass, \eq{9c}, still makes sense  
(when all events occur and fields are studied in a small part of 
the universe). Clearly, all these observations apply only to those 
matter distributions whose conformal size, $\sigma^1 \sim r^{pulse}$,
is small compared to $\pi$, the conformal size of the universe.
In other words, we will be interested in considering large wave numbers,
\begin{equation}
       n \sim \frac{1}{r^{pulse}} \gg 1
\label{16aa}
\end{equation}
%Also, the notion of localized pulses of matter is valid provided that the
%energy--momentum density contrast is large compared to the average energy
%density in one-dimensional universe,
%\begin{equation}
%     \frac{p_{\pm}^{pulse}}{r^{pulse}} \gg <p_{\pm}>
%\label{14*}
%\end{equation}
%where
%\[
%      p_{\pm}^{pulse} = \int_{pulse}~d\sigma^1~ \frac{1}{2\pi}
%                         (\partial_{\pm}{\bf f})^2
%\]
%and
%\[
%      <p_{\pm}> = \frac{1}{\pi}
%        \int_{-\pi/2}^{+\pi/2}~d\sigma^1~ \frac{1}{2\pi}
%                         (\partial_{\pm}{\bf f})^2
%\]
%In terms of the typical wave number $n$, \eq{14*} means
%\begin{equation}
%    p_{\pm}^{pulse}n \gg <p_{\pm}>
%\label{14+}
%\end{equation}
%This relation becomes non-trivial if, besides the pulses of matter, there 
%exists homogeneous matter distribution in the universe, 
%${\bf f}^{hom} \propto \sigma^0$. In what follows we always assume 
%eq. (\ref{14*}) to be valid.

The gauge (\ref{9+}) is not the only useful one in the closed universe.
At large $\gamma$, one can choose the gauge
\begin{equation}
     \phi' = -\frac{P^{(+)}}{\gamma} \sigma'^{0}
\label{12*}
\end{equation}
where prime is used to denote the quantities in this gauge. In general, 
the coordinates in the two gauges are not too different,
\begin{equation}
  \sigma'_{\pm} = \sigma_{\pm} + O(\gamma^{-1})
\label{17b}
\end{equation}
The coordinate transformation has a particularly simple form in the 
case of two narrow pulses; it follows from eqs.(\ref{13b}) and (\ref{14b})
that in that case
\begin{equation}
     \sigma'_{\pm} = \sigma_{\pm} + 
     \frac{\pi \mu}{\gamma P^{(+)}} |\sigma_{\pm}|
     - \frac{\mu}{\gamma P^{(+)}} \sigma_{\pm}^2 + O(\gamma^{-2})
\label{17d}
\end{equation}
Furthermore, for two narrow pulses one obtains (at $|\sigma_{\pm}|<\pi/2$)
\begin{equation}
     \rho'_{\pm} = -\frac{\pi \mu}{2\gamma P^{(+)}} \epsilon (\sigma'_{\pm})
\label{17c}
\end{equation}
This expression, of course, solves the constraint (\ref{8*}) for
matter distribution
\begin{equation}
     (\partial'_{\pm} {\bf f})^2 (\sigma'_{\pm}) = 
        \frac{\pi}{2} \mu \delta (\sigma'_{\pm})
\label{17a}
\end{equation}
The fact that the matter distributions (\ref{13a}) and (\ref{17a}) are 
essentially the same in the two gauges, is an immediate consequence 
of \eq{17b}.

If the pulses of matter are not infinitely narrow, the above features 
remain valid qualitatively. The matter distributions in the two gauges
(\ref{9+}) and (\ref{12*}) are the same up to corrections of 
order $\gamma^{-1}$; in particular, the spatial sizes of the lumps 
differ only by a factor $(1 + O(\gamma^{-1}))$. The scale factor
$\rho'(\sigma')$ changes rapidly in the regions of non-vanishing
energy--momentum density of matter, and has the form shown in fig.3,
the depth of the well being proportional to the total energy of mater.
We will have to say more about this gauge in section 5.

\section{Quantum states and vertex operators}
The model is quantized exactly in the same way as bosonic string theory
in Minkowskian target space of $D=26$ dimensions. One introduces the
notation
\[
     \phi=-\frac{1}{\gamma} (X^0 + X^1)
\]
\[
     \rho=-\frac{1}{\gamma} (X^0 - X^1)
\]
\begin{equation}
     f^1,\dots,f^{24} = X^2,\dots,X^{25}
\label{2.1+}
\end{equation}
Then $X^{\mu}(\sigma)$, $\mu=0,\dots,25$ are canonically normalized free
two-dimensional fields and the classical constraints (\ref{8*}) become
\begin{equation}
      \frac{1}{2} \partial_{\pm}X_{\mu} \partial_{\pm}X^{\mu}
           - \frac{1}{4} \gamma \partial^2_{\pm} X^{(+)} = 0 
\label{2.1*}
\end{equation}
where the summation is performed with Minkowskian $D$-dimensional
metrics, $\eta_{\mu \nu} = (-1,+1,\dots,+1)$ and 
\[
     X^{(\pm)} = X^0 \pm X^1
\]
Upon quantization, the left- and right-moving components of these 
fields are decomposed in the usual way,
\[
   X^{\mu}_{L}(\sigma_{+})=
         \frac{1}{2}x^{\mu} + \frac{1}{2}P^{\mu}\sigma_{+} +
         \frac{i}{2}\sum_{k\neq 0}\frac{1}{k}\alpha^{\mu}_{k}
         \mbox{e}^{-2ik\sigma_{+}} 
\]
\[
   X^{\mu}_{R}(\sigma_{-})=
         \frac{1}{2}x^{\mu} + \frac{1}{2}P^{\mu}\sigma_{-} +
         \frac{i}{2}\sum_{k\neq 0}\frac{1}{k}\tilde{\alpha}^{\mu}_{k}
         \mbox{e}^{-2ik\sigma_{-}} 
\]
where $\alpha_{k}^{\mu}$ and $\tilde{\alpha}_{k}^{\mu}$ are the standard
oscillator operators with string normalization. 
According to \eq{2.1*}, the Virasoro operators are
\[
   L_{0} = \frac{1}{8} P^{2} + \sum_{k>0} \alpha_{-k}^{\mu}
		       \alpha_{k}^{\mu}
\]
\begin{equation}
L_{m} = \frac{1}{2} P_{\mu} \alpha^{\mu}_{m} + 
         \frac{1}{2}\sum_{k\neq 0,m} :\alpha^{\mu}_{k}
	      \alpha^{\mu}_{m-k}: 
        + \frac{i}{2}\gamma m \alpha^{(+)}_{m}
\label{2.2*}
\end{equation}
and similarly for $\tilde{L}_0$ and $\tilde{L}_m$, where
$\alpha^{(\pm)}_{m} = \alpha^0_m \pm \alpha^1_{m}$. Equation (\ref{2.2*})
shows that the model is equivalent to the string with background 
charge\cite{Verlinde1,Verlinde2,Kazama1,Kazama2} or, in other words,
to the bosonic string in linear Dilaton background\cite{Tseytlin1,Tseytlin2}
\begin{equation}
    \Phi(x) = \gamma (e_{\mu}x^{\mu}) = \gamma x^{(+)}
\label{2.2+}
\end{equation}
with $e^{\mu} = (-1,1,0,\dots,0) $ being a light-like vector in
$D$-dimensional target space.

It is clear from \eq{2.2*} that the spectrum of states in the target space
is the same as the spectrum of the bosonic string in trivial background,
\[
     M^2 = 8(n-1),~~~~ n=0,1,\dots
\]
The vertex operators are, however, slightly different. For example, 
the tachyon vertex operator of conformal dimension (1,1) is
\begin{equation}
     V(Q) = : \e^{iQ_{\mu}X^{\mu} + \gamma X^{(+)}}:
\label{2.2**}
\end{equation}
with $Q^2 = 8$. This modification can be understood as being due to
the Dilaton background (\ref{2.2+}). Indeed, the effective action of 
tachyon field $\tilde{T}$ in flat target space-time and in linear Dilaton
background $\Phi (x)$ is
\[
     \int~d^D x~ \e^{-2\Phi} \left[ -\frac{1}{2} (\partial_{\mu} \tilde{T})^2
                + 4 \tilde{T}^2 + c\tilde{T}^3 + O(\tilde{T}^4)\right]
\]
By introducing the field $T= \e^{-\Phi} \tilde{T}$ one rewrites this 
action in the following form,
\[
     \int~d^D x~  \left[ -\frac{1}{2} (\partial_{\mu} T)^2
                - \frac{M_T}{2} T^2 + c \e^{\Phi} T^3 + O(T^4) \right]
\]
where $M_T^2 = -8 + (\partial_{\mu}\Phi)^2$, i.e., $M_T^2 = -8$ for 
light-like $e_{\mu}$ in \eq{2.2+}. In this notation
the kinetic and mass terms are conventional, while the
 trilinear vertex is proportional to $\e^{\Phi}$, precisely as required 
by \eq{2.2**}. This argument can be generalized to interactions of the fields
other than the tachyon\cite{Tseytlin1,Tseytlin2,TseytlinPrivate}.

Let us now consider highly excited string states (parent universes). 
They are conveniently constructed by making use of the DDF operators\cite{DDF}.
In the 
light-like Dilaton background, the simplest choice of the DDF operators is
\[
   A^{i}_{n}=\int\limits_{-\pi/2}^{+\pi/2}~\frac{d\sigma_{+}}{\pi}
              \exp\left[4in\frac{e_{\mu}
          X^{\mu}_{L}(\sigma_{+})}{e_{\mu}P^{\mu}}\right]
	     \partial_{+}X^{i}_{L}(\sigma_{+})
\]
\begin{equation}
   \tilde{A}^{i}_{\tilde{n}}
   =\int\limits_{-\pi/2}^{+\pi/2}~\frac{d\sigma_{-}}{\pi}
              \exp\left[4i\tilde{n}\frac{e_{\mu}
          X^{\mu}_{R}(\sigma_{-})}{e_{\mu}P^{\mu}}\right]
	     \partial_{-}X^{i}_{R}(\sigma_{-})
\label{2.4*}
\end{equation}
with the same light-like vector $e_{\mu}$.
Here $i=2,\dots,25$. These operators obey the usual oscillator commutational
relations, and their commutational relations with the 
Virasoro operators\cite{VR}
ensure that the state of the form
\begin{equation}
      A_{-n_1}^{i_1} \dots A_{-n_s}^{i_s}\cdot
          \tilde{A}_{-\tilde{n}_1}^{j_1} \dots \tilde{A}_{-\tilde{n}_t}^{j_t}
      \ket{{\cal P}}
\label{2.4+}
\end{equation}
is a physical state provided that $\ket{{\cal P}}$ is the physical 
tachyon state and
\begin{equation}
    n_1 + \dots + n_s = \tilde{n}_1 + \dots + \tilde{n}_t
\label{2.4+++}
\end{equation}
The DDF operators (\ref{2.4*}) are similar to those introduced in 
ref.\cite{Kazama2}. Another choice of DDF operators in this model
has been considered in refs.\cite{Verlinde2,Kazama2}.

To make contact with the classical analysis of section 2, consider suitably 
modified coherent states
\begin{equation}
    \ket{\Psi_P} =
       P_{L_0=\tilde{L}_0}
       \exp \left( \sum_{n>0} \frac{1}{n} f^i_n A^i_{-n} +
             \sum_{\tilde{n}>0} \frac{1}{\tilde{n}} \tilde{f}^j_{\tilde{n}}
              \tilde{A}^j_{-\tilde{n}}\right)
         \ket{{\cal P}}
\label{2.5+}
\end{equation}
Here $f_n^i$ and $\tilde{f}^j_{\tilde{n}}$ are $c$-number amplitudes and
 $P_{L_0=\tilde{L}_0}$ is a projector onto the subspace of vectors obeying
\[
       (L_0 - \tilde{L}_0) \ket{\Psi} = 0
\]
This projector is needed to ensure the validity of \eq{2.4+++} term by term in
the expansion of $\ket{\Psi_P}$.

Let us impose the condition that
\begin{equation}
     \sum_{n} {\bf f}_n {\bf f}^{*}_n
               = \sum_{\tilde{n}} \tilde{{\bf f}}_{\tilde{n}}
           \tilde{{\bf f}}^{*}_{\tilde{n}}
\label{2.5**}
\end{equation}
and take the amplitudes ${\bf f}_n$ and $\tilde{{\bf f}}_{\tilde{n}}$ to be 
large.  Consider now matrix elements of the form
\begin{equation}
       \bra{\Psi_{P'}} O_{M}(\{ \sigma_{+} \};\{ \sigma_{-} \})
        \ket{\Psi_P}
\label{2.5*}
\end{equation}
and also
\begin{equation}
\bra{\Psi_{P'}} \partial_{-} X_{R}^{(\pm)} (\sigma'_{-})
           O_{M}(\{ \sigma_{+} \};\{ \sigma_{-} \})
        \ket{\Psi_P}
\label{++++}
\end{equation}
\begin{equation}
 \bra{\Psi_{P'}} \partial_{+} X_{L}^{(\pm)} (\sigma'_{+})
           O_{M}(\{ \sigma_{+} \};\{ \sigma_{-} \})
        \ket{\Psi_P}
\label{2.6+}
\end{equation}
where the operators $O_M$ are products of matter fields, in general 
at different points in (1+1)-dimensional space-time,
\begin{equation}
 O_{M} = \partial_{+}X_{L}^{i_1}(\sigma^1_{+})\dots
          \partial_{+}X_{L}^{i_k}(\sigma^{k}_{+})\cdot
\partial_{-}X_{R}^{j_1}(\sigma^1_{-})\dots
          \partial_{-}X_{R}^{j_q}(\sigma^{q}_{-})
\label{2.6**}
\end{equation}
and, as before, we take $i_1,\dots,j_q = 2,\dots,25$, so that the 
operators $O_M$ indeed contain matter fields only. The matrix elements
(\ref{2.5*}) are then the correlators of matter fields in the coherent
state (\ref{2.5+}), while the matrix elements (\ref{++++}) and
(\ref{2.6+}) are the
correlators of dilaton and metric fields with matter.

These matrix elements are calculated in Appendix with the following
result. Up to small corrections and trivial normalization
factor, they coincide with classical correlators
\begin{equation}
     \int_{-\pi/2}^{+\pi/2} ~\frac{d\xi^1}{\pi} ~
       O_{M}^{cl}(\{ \sigma_{+} + \xi^1 \}; \{ \sigma_{-} - \xi^1 \} ) \equiv
 \int_{-\pi/2}^{+\pi/2}~ \frac{d\xi^1}{\pi}~ 
       O_{M}^{cl}(\{ \sigma^{0} \};\{ \sigma^1 + \xi^1 \} )
\label{2.7*}
\end{equation}
and
\begin{equation}
\int_{-\pi/2}^{+\pi/2} ~\frac{d\xi^1}{\pi} ~
       \partial_{+} X^{cl, (\pm)}_{L} (\sigma_{+} + \xi^1)
       O_{M}^{cl}(\{ \sigma_{+} + \xi^1 \}; \{ \sigma_{-} - \xi^1 \} )
\label{2.7+}
\end{equation}
(and similarly for \eq{++++})
respectively, where $O_M^{cl}$ is given by \eq{2.6**} with classical 
matter fields
\[
  \partial_{+} X^{cl,i} (\sigma_{+}) =
         \frac{1}{2} P^i + \sum_{n>0} \left( f_n^i \e^{-2in\sigma_{+}}
              + f_n^{*i} \e^{2in\sigma_{+}} \right)
\]
The classical field $X^{cl,(+)}$ is defined by
\[
       \partial_{\pm} X^{cl,(+)} = {\cal P^{(+)}}
\]
while the field $\partial_{\pm} X^{cl,(-)}$ is to be found from the
classical constraints (\ref{2.1*}). In short, the matrix elements like 
\eq{2.5*} and \eq{2.6+} are equal to the corresponding classical 
expressions {\it in the gauge} (\ref{12*}), integrated over translations in
one-dimensional space. In this way the classical picture is restored; 
the particular choice of the DDF operators, \eq{2.4*}, corresponds to the
gauge choice (\ref{12*}). 
Note that \eq{2.5**} is precisely the classical constraint
$L_0 = \tilde{L}_0$ written in this gauge.

Hence, the DDF operators (\ref{2.4*}) correspond to creation and 
annihilation of dressed matter excitations in (1+1)-dimensional space-time
{\it in the gauge} $\phi = {\rm const}\cdot\sigma^0$. 
The corresponding wave numbers are
$n$ and $\tilde{n}$, respectively. In what follows we will call (somewhat
loosely) these excitations as ``dressed particles'' in the (1+1)-dimensional
universe.

\section{Non-conservation of bare energy in (1+1) dimensions and emission 
          probability}
\subsection{State of the parent universe}
In this section we consider the simplest DDF state
\begin{equation}
   \ket{\Psi, n, i,j}=
         \int~d^{D-1} P~ \Psi(P) \ket{P,n,i,j}
\label{3.1*}
\end{equation}
where 
\begin{equation}
      \ket{P,n,i,j} = \frac{1}{n} A_{-n}^i \tilde{A}_{-n}^j 
                             \ket{{\cal P}}
\label{3.1+}
\end{equation}
and $\Psi(P)$ is the wave function of the center-of-mass motion
in target space, in
momentum representation. The normalization factor $1/n$ in \eq{3.1+} is
chosen in such a way that the state $\ket{P,n,i,j}$ has the usual
$D$-dimensional normalization (recall  the string normalization
of the oscillator operators, $[A^i_n,A^j_n] = n \delta^{ij}$).
According to the discussion in section 3, we interprete this state as
the state of a parent universe with two dressed
matter particles (one left-moving and
one right-moving) of equal wave numbers $n$, in the gauge (\ref{12*}). The
normalization convention in \eq{3.1+} corresponds to ``two particles in 
entire one-dimensional space'' normalization of (1+1)-dimensional quantum 
field theory. These particles have equal bare energies and opposite 
bare momenta in (1+1) dimensions,
\begin{equation}
       \epsilon_{left} = \epsilon_{right} =  2n
\label{3.1**}
\end{equation}
\[
       p_{right} = - p_{left} = 2n
\]
We are interested in the limit (see \eq{16aa})
\[
     n \to \infty
\]
By making superpositions of the states (\ref{3.1*}) with different 
wave numbers $n$  one can construct states with localized distributions
of matter in one-dimensional universe. 
This generalization is straightforward, so we stick to the
state with fixed $n$. The necessity to consider wave packets (\ref{3.1*})
in target space
and not just plane waves (\ref{3.1+}) is due to the Dilaton
background that increases indefinitely as $x^{(+)} \to \infty$:
amplitudes of processes involving plane waves would be divergent 
in this background.

It is a matter of simple algebra to see that the total $D$-dimensional 
momentum of the state (\ref{3.1+}) is
\[
          P^k = {\cal P}^k,~~~ k = 2,\dots,25
\]
\[
          P^{(+)} \equiv P^0 + P^1 = {\cal P}^{(+)}
\]
\begin{equation}
          P^{(-)} \equiv P^0 - P^1 = 
              {\cal P}^{(-)} - \frac{8n}{{\cal P}^{(+)}}
\label{3.2+}
\end{equation}
Hence, from $D$-dimensional point of view, the state (\ref{3.1+})
is interpreted as an excited string state at the $n$-th level
with the mass
\begin{equation}
     M_n^2 = 8n-8
\label{3.2**}
\end{equation}
(recall that $\ket{{\cal P}}$ is the tachyon state). We consider for 
definiteness this state in the center-of-mass frame,
\[
    P^k = 0,~~~ k=2,\dots,25
\]
\[
    P^{(+)} = {\cal P}^{(+)} = M_n
\]
\begin{equation}
    P^{(-)} = M_n
\label{3.2*}
\end{equation}
although our discussion can be straightforwardly generalized to
other frames. In terms of the wave packets (\ref{3.1*}), \eq{3.2*}
means that the wave function of the center-of-mass motion,
$\ket{\Psi (P)}$ is peaked near the values determined by \eq{3.2*}.
%Note that due to \eq{3.2*} the {\em average} (1+1)-dimensional energy
%also coincides with $2\epsilon_n$ and is of order $n$, so that the
%relation (\ref{14+}) indeed holds at large $n$. In other words, 
%appropriate superpositions of the states (\ref{3.1*}) indeed describe
%localized states of matter in (1+1) dimensions.

\subsection{Emission of baby universe: energy non-conservation
            in (1+1) dimensions}
The parent universe in the state (\ref{3.1*}) can emit a baby universe ---
a universe with no or small energy of matter particles. In $D$-dimensional
language this process corresponds to the emission of a low lying string state
(tachyon, Dilaton, etc.) into $D$-dimensional target space.

Let us study whether this process always occurs with non-conservation of
bare energy in (1+1) dimensions. Since bare energy coincides with the 
level of the string, \eq{3.1**}, we are interested in the change of the 
$D$-dimensional mass of the highly excited string due to the emission of
the low lying string state. At first sight, the presence of the
time-dependent Dilaton background in target space might give rise to 
non-conservation of $D$-dimensional energy and momentum in the emission
process. In that case the emission of a low lying state would not
necessarily require the change of the level of the heavy string; in 
other words, the emission of a baby universe would not necessarily require
non-conservation of bare energy in (1+1) dimensions.
Surprisingly, we will see in a moment that this is not the case: 
$D$-dimensional energy and momentum are in fact conserved in the
presence of {\it linear} Dilaton background exactly as they do in 
flat target space with no background. Hence, the emission of a baby 
universe always occurs with non-conservation of (1+1)-dimensional
bare energy of matter in the parent universe.

The argument presented below is fairly general; the particular form of
the initial state, \eq{3.1*}, is unimportant. Non-conservation of
energy in (1+1)-dimensional parent universe due to the emission of
baby universes is a generic property of our model.

The argument for the {\it conservation} of $D$-dimensional
energy--momentum in the presence of the linear Dilaton background
is conveniently presented by considering  a theory
of three scalar fields $\Phi_1$, $\Phi_2$, $\Phi_3$ with
cubic interaction and
exponentially changing coupling.
Let the cubic 
coupling  be
\begin{equation}
        g(x) = \exp (\Gamma_{\mu} x^{\mu})
\label{3.5*}
\end{equation}
with real constant $\Gamma_{\mu}$.
Let us take $\Phi_1$ to be the heaviest (mass $M_i$), and
$\Phi_2$ to have the mass $M_f$ which for simplicity is close to
$M_i$ (but smaller than $M_i$). Let the field $\Phi_3$ be light.
We are interested in the process when a particle $\Phi_1$ emits a
particle $\Phi_3$ and becomes a particle $\Phi_2$.

Because of the unboundedness of the coupling (\ref{3.5*}), it does not 
make sense to consider plane waves of particles $\Phi_1$ and $\Phi_2$. 
Instead, one has to use wave packets. Namely, consider the amplitude
of the decay of a wave packet
$\Psi_i ({\bf x},t)$ describing the state of the particle $\Phi_1$ 
into a wave packet $\Psi_f$ of the particle $\Phi_2$ plus a
particle $\Phi_3$ which has fixed energy--momentum $Q_{\mu}$,
\begin{equation}
   A =
         \int~d{\bf x}~dt~ g({\bf x},t)
         \Psi_i ({\bf x},t) \Psi_f^{*} ({\bf x}, t)
         \exp(-i Q_{\mu} x^{\mu})
\label{3.6*}
\end{equation}
We will see in the next subsection that the amplitudes of string decays
have similar form.

Let us specify the form of the wave packets $\Psi_i$ and $\Psi_f$. Let us
consider for definiteness the non-relativistic regime and take wave packets
narrow in momentum representation. Let us furthermore neglect the dispersion
of the wave packets with time\footnote{The dispersion of wave
packets with time would complicate the analysis considerably.
The coupling changes so rapidly
that in the case of spreading wave packets the interaction often occurs
in space-time very
far away from the centers of the  wave packets. In that region
there are mostly modes
with momenta quite different from the central values ${\bf P}_i$
or ${\bf P}_f$.  Hence it is difficult to
separate the effects of finite widths of the wave packets in momentum
space from possible effects of momentum non-conservation.} (this can be 
achieved by confining  the particles in moving potential wells). Thus, we take 
\begin{equation}
   \Psi_i = \exp \left( -iM_i t - \frac{i}{2M_i}{\bf P}_i^2
   + i {\bf P}_i {\bf x} - \frac{\sigma^2}{2} ({\bf x} - {\bf v}_i t)^2
   \right)
\label{3.7*}
\end{equation}
where $\sigma$, the width of the packet in momentum space, is small,
and ${\bf v}_i = {\bf P}_i/M_i$ is the velocity of the particle, which is also
assumed to be small  for simplicity.  Similarly, the wave packet $\Psi_f$ 
has the form (\ref{3.7*}) with $M_i$, ${\bf P}_i$, ${\bf v}_i$
substituted by $M_f$, ${\bf P}_f$, ${\bf v}_f$.

The integral (\ref{3.6*}) is then Gaussian, and is straightforward to evaluate.
The result, up to a pre-exponential factor, is
\begin{equation}
    A = \exp \left[ \frac{1}{2\sigma^{2}}
                   \left(- F_1({\bf \Delta P}, \Delta E)
                 + F_1({\bf \Gamma}, \Gamma_0)
                 + i F_2 ({\bf \Delta P, \Gamma}, \Delta E, \Gamma_0)
                  \right) \right] 
\label{3.8*}
\end{equation}
where
\[
       {\bf \Delta P} = {\bf P}_i - {\bf P}_f - {\bf Q}
\]
\[
       \Delta E = \left( M_i + \frac{{\bf P}_i^2}{2 M_i} \right)
                - \left( M_f + \frac{{\bf P}_f^2}{2 M_f} \right)
                - Q_0   
\]
are amounts of non-conservation of  momentum and energy and
\[
    F_1 ({\bf \Delta P}, \Delta E) =
       \frac{({\bf \Delta P})^2}{4} +
        \frac{1}{({\bf v}_i  - {\bf v}_f)^2}
         \left[ \Delta E - 
          \frac{1}{2} ( {\bf v}_i - {\bf v}_f) {\bf \Delta P}
          \right]^2
\]
\[
    F_1 ({\bf \Gamma}, \Gamma^0) =
       \frac{({\bf \Gamma})^2}{4} +
        \frac{1}{({\bf v}_i  - {\bf v}_f)^2}
         \left[ \Gamma^0 - 
          \frac{1}{2} ( {\bf v}_i - {\bf v}_f) {\bf \Gamma}
          \right]^2
\]
The explicit form of $F_2$ is not important; it is sufficient to
note that both $F_1$ and $F_2$ are real.
The only important property of $F_1$ is that it is positive-definite
and vanishes iff ${\bf \Delta P} = \Delta E = 0$.

The imaginary part of the exponent, $i F_2$, in \eq{3.8*} 
is unimportant and
cancels out in the probability. Then the probability factorizes
into a term depending on ${\bf \Delta P}$ and $\Delta E$ and a
term containing $\Gamma_{\mu}$. The latter term is nothing but the overlap
of the initial and final wave packets with the weight $g(x)$.
More importantly, since $F_1 ({\bf \Delta P}, \Delta E)$ is multiplied by
the large factor $1/\sigma^2$ in the exponent, and because
$F_1({\bf \Delta P}, \Delta E)$ is non-negative and vanishes 
only at ${\bf \Delta P} = \Delta E = 0$, the exponential factor containing
$F_1({\bf \Delta P}, \Delta E)$ ensures conservation of energy and momentum in the limit 
of small $\sigma$ exactly in the same manner as it does in the case
of space-time independent coupling.

The restriction to the non-relativistic case and Gaussian wave packets
is, in fact, not essential:
the same argument goes through for relativistic and
non-Gaussian wave packets (provided
that they do not disperse with time). 

The result that energy and momentum are
conserved in spite of space-time dependence of the coupling
is peculiar to the exponential coupling whose
exponent linearly depends on $x_{\mu}$: only in this case the
dependences on $\Delta P_{\mu}$ and $\Gamma_{\mu}$ factorize.
For instance, if the coupling switches off at infinity  in
space-time (say, has finite support), then the same calculation
leads to the usual result that energy and momentum are not conserved,
and the amplitude is proportional to the Fourier component  of the coupling,
$\tilde{g}({\bf \Delta P}, \Delta E)$.

\subsection{Emission rate}
Let us now turn to the actual calculation of the rate of the
emission of a baby universe (low-lying string state) by a parent  universe
(excited string) in the state (\ref{3.1*}). We explicitly consider the 
emission of a tachyon, although the analysis --- and results --- are 
the same for the emission of a
Dilaton or graviton as the baby universe. Let the
$D$-dimensional momentum of the outgoing tachyon be $Q_{\mu}$. We first 
have to specify the range of $Q_{\mu}$ which is of interest for our purposes.

The emission of a baby universe due to  the collision of
(1+1)-dimensional particles
with wave numbers equal to 
$n$ has a chance to be local if the characteristic 
conformal time of the process of emission, $\Delta \sigma^0$, is of order
\[
 \Delta \sigma^0 \sim \frac{1}{n}
\]
This conformal time is related to the $D$-dimensional
time via
\begin{equation}
   \Delta x^0 = P^0 \Delta \sigma^0
\label{3.11*}
\end{equation}
The $D$-dimensional time characteristic to the tachyon emission
can be estimated in the center-of-mass frame of the decaying string as
\[
   \Delta x^0 \sim \frac{1}{Q^0}
\]
This is the time after which the tachyon is formed and splits off the
initial string. Hence we are interested in the tachyon energies
of order
\begin{equation}
     Q^0 \sim \frac{n}{P^0} \sim \sqrt{n}
\label{3.12*}
\end{equation}
where we made use of eqs.(\ref{3.2**}) and (\ref{3.2*}). Since
$Q^0$ is large,  we neglect the tachyon mass where appropriate
in what follows.

We are interested in the process in which the initial string
at level  $n$  decays, by emitting a tachyon, into level $n'$.
According to eqs.(\ref{3.2**}) and (\ref{3.12*}), and because of 
energy conservation in $D$ dimensions, we have
\begin{equation}
           m\equiv n-n' \sim n
\label{3.12+}
\end{equation}
We intend to sum up over all final states at given level $n'$. Usually, 
this summation is conveniently performed by evaluating the forward 
amplitude shown in fig.4 where curved lines denote the tachyon 
(cf.  ref. \cite{Nirov}). Because of the space-time
dependent coupling, the procedure is somewhat tricky in our case.
It will be convenient to consider first the amplitude with different 
momenta, as shown in fig.5, and set $P=P'$ and $K=Q$ in  the end.
The amplitude is given by
\begin{equation}
    A^{ij}(P,P';K,Q) =
         \kappa^2 \bra{P',n,i,j} V(K) \Delta V(Q) \ket{P,n,i,j}
\label{3.13*}
\end{equation}
where $\kappa$ is the string coupling constant, $V$ is the vertex operator 
(\ref{2.2**}) and $\Delta$ is the usual string propagator
(recall that the Virasoro operators
$L_0$ and $\tilde{L}_0$ coincide with conventional ones). Here we consider 
the plane wave (\ref{3.1+}) as the initial state; the fact that we actually
have to deal with wave packets like (\ref{3.1*}) has been already discussed
in the previous subsection\footnote{The procedure below involves 
manipulations with formally divergent integrals, etc. This procedure
can be checked by explicit analysis of amplitudes involving low
lying string states only.}. 

The evaluation of the amplitude (\ref{3.13*}) is straightforward and
parallels that of ref.\cite{Nirov}. One finds
\begin{eqnarray}
    A^{ij}(P,P';K,Q) &=& \kappa^2 \int~d^D x~ \e^{i(P-Q)x}
                       \e^{-i(P'-K))x} \e^{2\gamma x^{(+)}} \nonumber \\
&& \nonumber \\
&&           \times~  \int~dz d\bar{z} ~
     (z \bar{z})^{\frac{1}{4}(P^{\mu} - i\gamma e^{\mu})
        (-Q_{\mu} - i\gamma e_{\mu})} \frac{1}{n^2} F^{ij}(z, \bar{z})
\label{3.14*}
\end{eqnarray}
where we have kept the integration over the center-of-mass coordinate
$x$ and denoted
\begin{equation}
   F^{ij}(z, \bar{z}) = |1-z|^{\frac{1}{2}(K^{\mu} - i\gamma e^{\mu})
        (-Q_{\mu} - i\gamma e_{\mu})}
          B^{ij}(z)B^{ij}(\bar{z})
\label{3.14**}
\end{equation}
with
\begin{eqnarray}
B^{ij}(z) &=&
                    \int~\frac{du}{2\pi}
                    \frac{du'}{2\pi} \frac{1}{u^{n+1}}
\frac{1}{(u')^{n+1}} \nonumber \\
&& \nonumber \\
&&  \times~ \left( 1-u \right)^{-\frac{nK^{(+)}}{P^{(+)}}}
                     \left( 1-\frac{u}{z} \right)^{\frac{nQ^{(+)}}{P^{(+)}}}
     \left( 1-u'\right)^{\frac{nK^{(+)}}{P'^{(+)}}} 
      \left( 1-zu' \right)^{-\frac{nQ^{(+)}}{P'^{(+)}}} \nonumber \\
&& \nonumber \\
&& \times~  \left[ \frac{1}{4}
        \left( P^{i} + Q^i\frac{u}{z-u} - K^i \frac{u}{1-u}\right)
 \left( P'^{j} + K^j\frac{u'}{1-u'} - Q^j \frac{zu'}{1-zu'}\right) \right.
\nonumber \\
&& \nonumber \\
&& \left.        +~ \delta^{ij} \frac{uu'}{(1-uu')^2} \right]
\label{3.14+}
\end{eqnarray}
The integrations here run around small circles in complex $u$- and $u'$-planes
surrounding the origin. In eqs.(\ref{3.14*}) and (\ref{3.14**}) we used the
light-like vector $e_{\mu} = (-1,1,0,\dots,0)$ defined in section 3.

To extract the decay rate into final states at given level $n'$, we
expand $F(z, \bar{z})$ in a formal series in $z$, $\bar{z}$ (we
omit superscripts $i$, $j$ temporarily),
\begin{equation}
   F(z,\bar{z})= \sum_{m,m'=-\infty}^{+\infty} F_{m m'}z^{-m} \bar{z}^{-m'}
\label{3.15*}
\end{equation}
At $m=m'$ the corresponding integrals in \eq{3.14**} have poles,
\begin{eqnarray}
   F_{m m} \int~dz d\bar{z} &~& 
     (z \bar{z})^{\frac{1}{4}(P^{\mu}  i\gamma e^{\mu})
        (-Q_{\mu} - i\gamma e_{\mu}) -m} \nonumber \\
&& \nonumber \\
&& \sim
         \frac{\pi F_{m m}}
     {-\frac{1}{4}(P^{\mu} - i\gamma e^{\mu})(-Q_{\mu} - i\gamma e_{\mu})
     -m-1} \nonumber \\
&& \nonumber \\
&&   \sim \frac{8\pi F_{m m}}{(P_{\mu} - Q_{\mu} - i\gamma e_{\mu})^2 
        + (M_n^2 - 8m)}
\end{eqnarray}
These pole terms lead to contributions 
to the amplitude (\ref{3.14*}) which can be written in the following form,
\begin{eqnarray}
   A^{ij}(P,P';K,Q) &=& \sum_{m} 8\pi \kappa^2 \frac{1}{n^2} F_{m m}
        \int~d^Dx~d^D y d^D P_f~
          \e^{i(P-Q)x + \gamma x^{(+)}} \nonumber \\
&& \nonumber \\
&&         \frac{\e^{-iP_f (x-y)}}{(2\pi)^D (-P_f^2 - M_{n-m}^2)}
       \e^{-i(P'-K)y + \gamma y^{(+)}}
\end{eqnarray}
This expression is recognized as the sum of the amplitudes of processes
going through states with masses $M_{n-m}$ in a theory with trilinear 
coupling which exponentially depends on $x^{(+)}$. Hence, the probability 
of the decay into level $n' = n-m$ is determined by $F_{m m}$. The
total probability involves also the overlap between the initial and final
wave functions of the center-of-mass motion. This overlap has been 
considered in previous subsection; it leads to the conservation of energy and
momentum in $D$ dimensions. Therefore, we can set $P'=P$ and $K=Q$ and obtain
\begin{equation}
 \sum_{f} |A^{ij}_{f} (n \to n-m;Q)|^2 = 8 \pi \kappa^2 \frac{1}{n^2}
                    F_{m m}^{ij} (P'=P; K=Q)
\label{3.16ba}
\end{equation}
where $A^{ij}_f$  denotes the amplitude of the decay into a final state $f$ 
at level $(n-m)$ and a tachyon with momentum $Q$; the sum in the left hand side
runs over all final states at the level $(n-m)$. This expression should be
integrated over the phase space of the two final string states.

To estimate the integral over the phase space, let us study first the 
behavior of $F_{m m}$ at large $n$ and $m$. We recall that we are 
interested in tachyons with energies $Q^0 \sim \sqrt{n}$.  Let us first 
consider the generic case,
\begin{equation}
      Q^{(+)} \sim Q^{(-)} \sim Q^{k} \sim \sqrt{n}
\label{3.12**}
\end{equation}
At $K=Q$ and $P'=P$ we have
\begin{equation}
       F^{ij}_{mm'} = R^{ij}_m R^{ij}_{m'}
\label{3.17**}
\end{equation}
where the form of $R_m^{ij}`$ follows from eqs.(\ref{3.14**}),
(\ref{3.14+}) and (\ref{3.15*}),
\begin{eqnarray}
R^{ij}_m &=&
                    \int~\frac{du}{2\pi}
                    \frac{du'}{2\pi} \frac{dz}{2\pi}~
     \frac{1}{u^{n+1}(u')^{n+1} z^{-m+1}} \nonumber \\
&& \nonumber \\
&&  \times~ (1-z)^{-2} 
   \left[ \frac{\left(1-\frac{u}{z} \right)\left( 1-u'\right)}{(1-u)(1-zu')}
    \right]^{\frac{nQ^{(+)}}{P^{(+)}}}  \nonumber \\
&& \nonumber \\
&& \times~  \left[ \frac{1}{4}
        \left( P^{i} + Q^i\frac{u(1-z)}{(z-u)(1-u)}\right)
     \left( P^{j} + Q^j\frac{u'(1-z)}{(1-u')(1-zu')}\right) \right.\nonumber \\
&&\nonumber \\
&&  \left.  + ~\delta^{ij} \frac{uu'}{(1-uu')^2} \right]
\label{3.17+}
\end{eqnarray}
In the regime (\ref{3.12**}), $R_m$ can be written as follows,
\begin{equation}
    R_m^{ij} = \int ~du~du'~dz~ P^{ij}(u,u',z) \e^{-nS(u,u',z)}
\label{3.17*}
\end{equation}
where
\[
  S = \log{u} + \log{u'} - \frac{m}{n} \log{z}
       - \frac{Q^{(+)}}{M_n} \left[ -\log({1-u}) + \log({1-\frac{u}{z}})
        + \log({1-u'}) - \log({1-zu'}) \right]
\]
and $P^{ij}$ is a pre-exponential factor that depends on $n$
only weakly. Here we made use of  the fact that
    $P^{(+)} = M_n$
in the center-of-mass frame we consider.

In the regime (\ref{3.12**}) all terms in $S$ are of order one, so the
integral in \eq{3.17*} can be calculated by saddle point
technique. At the saddle point one finds
\begin{equation}
    S= \chi\left( \frac{m}{n}\right) - 
       \chi\left( \frac{m}{n} - \frac{2Q^{(+)}}{M_n} \right)
\label{3.18*}
\end{equation}
where
\[
   \chi(\nu) = (1-\nu) \log({1-\nu}) +(1+\nu) \log({1+\nu})
\]
Making use of energy-momentum conservation, one obtains (again in the
center-of-mass frame of the decaying string)
\[
     -\frac{m}{n} < 
       \left( \frac{m}{n} - \frac{2Q^{(+)}}{M_n} \right) < \frac{m}{n}
\]
so $S$ is always positive. Therefore, we conclude that the  decay 
probability is exponentially {\it suppressed} in the kinematical region
(\ref{3.12**}),
\[
  P(n \to n-m;Q) \propto \e^{-2nS}
\]

Equation (\ref{3.18*}) implies that the decay probability may be
{\it unsuppressed} in the kinematical region different from \eq{3.12**},
namely, at
\begin{equation}
     Q^{(+)} \sim \frac{1}{\sqrt{n}}
\label{3.19**}
\end{equation}
i.e., at
\begin{equation}
    \frac{nQ^{(+)}}{M_n} \sim 1
\label{3.19*}
\end{equation}
In this region
\[
      nS = \chi'\left(\frac{m}{n} \right) \cdot 2\frac{nQ^{(+)}}{M_n} \sim 1
\]
Clearly, the saddle point calculation is not valid in this region, so we 
proceed in a different way. To estimate the integral in \eq{3.17+}
in the regime (\ref{3.19*}), we make use of the following asymptotic
formula\cite{Nirov}
\[
  \int~dz_{1}\dots dz_{k} z_{1}^{-\lambda a_{1}}\dots
              z_{k}^{-\lambda a_{k}}
               \Pi (1-z_{p})^{\alpha_{p}}
               \Pi (1-z_{p}z_{q})^{\beta_{pq}}
               \Pi (1-z_{p}/z_{q})^{\gamma_{pq}}~
\]
\[
               \propto~
               \left(\frac{1}{\lambda}\right)^{\sum \alpha_{p} +
                 \sum \beta_{pq} + \sum \gamma_{pq} + k} \,,
\]
which is valid as $\lambda \to  \infty$ with $a_{p},\alpha_{p},
\beta_{pq}, \gamma_{pq}$ fixed. We obtain
\[
    R^{ij}_m \propto n
\]
at large $n$. We then recall eqs.(\ref{3.16ba}) and (\ref{3.17**}) and
find that at large $n$
\[
    \sum_{f} |A^{ij}_{f} (n \to n-m;Q)|^2 = {\rm~independent~ of~}n
\]
in the kinematical region (\ref{3.19**}).

To estimate the total emission probability, we note that in the region
(\ref{3.19**}) one has
\[
    {\bf Q}^2_{T} \sim  1
\]
where ${\bf Q}_{T}= (0,0,Q^2,\dots,Q^{25})$. Hence,  the probability 
to decay into the level $(n-m)$ is of order
\[
  P(n \to n-m) = \int~ \frac{d^{D-2} Q_{T}}{(2\pi)^{D-2}}~
                 \frac{1}{Q^0 M_n E_{n-m}}\sum_{f} |A^{ij}_{f} (n \to n-m;Q)|^2
                 \sim \frac{1}{n^{3/2}}
\]
where $E_{n-m} \sim M_n \sim \sqrt{n}$ is the energy of the final 
excited string, and $Q^0 \sim \sqrt{n}$ according to \eq{3.12*}.
The number of final states contributing to the decay, $\Delta m$, is of 
order $n$, so we have finally
\[
        \sum_{m} P(n \to n-m) \sim \frac{1}{\sqrt{n}}
\]
This is the decay probability  per unit $D$-{\it dimensional} time
$x^0$. To  obtain its interpretation in (1+1)-dimensional terms,
we recall \eq{3.11*} and find that  the emission rate  per unit 
conformal time $\sigma^0$ of (1+1)-dimensional universe is {\it 
independent} of $n$  at large $n$.

This is our  principal result: the rate of the emission
of baby universes is unsuppressed at large $n$, when the emission process 
should occur locally. This emission rate is proportional to the collision
rate of two narrow wave packets in one-dimensional universe of conformal
size $\pi$, the proportionality constant being independent of the 
one-dimensional momenta of the ``particles'' or width of their wave packets
and being determined by the string coupling constant only.

\section{Discussion and conclusion}
Let us summarize our results for the simplest version of the dilaton
gravity with conformal matter in (1+1) dimensions. We considered mostly
the case of compact one-dimensional universe and studied pulses of matter
whose size is small compared to the size of the universe (i.e.,
whose wave numbers $n$ are large). At least at the 
classical level these pulses, in the gauge
\begin{equation}
      \rho = {\rm const} \cdot \sigma^0
\label{4.1*}
\end{equation}
produce long ranged dilaton field which
is approximately Coulomb at scales small compared to the size of the universe.
The magnitude of this long ranged field is proportional to  the  energy
of matter, which we called bare energy. The notion of ADM mass makes 
sense at these scales and coincides with the ADM mass defined
for infinite space.

To construct quantum states, it was convenient to work in a different
gauge,
\begin{equation}
       \phi = {\rm const} \cdot \sigma^0
\label{4.2*}
\end{equation}
It is important that at large $\gamma$, the sizes of matter pulses 
in the two gauges are similar, again at the classical level. In other words,
the pulses that are narrow in the gauge (\ref{4.1*}) are also narrow in 
the gauge (\ref{4.2*}), so the processes we were interested in are 
local in both gauges. Making use of  the gauge (\ref{4.2*}), we considered ---
at the quantum level --- the simplest state of the parent universe 
that contains one left-moving and one right-moving dressed 
matter ``particles'' with large wave number $n$. The collisions of these 
particles may eventually induce the emission of baby universes.
If the relevant quantum numbers of the baby universe ($D$-dimensional
momenta $Q_{\mu}$ of the microscopic string) are large enough, 
the emission process is local in the parent universe. We have found that the
emission always occurs with non-conservation of energy of matter, and
that the probability of this process is finite at large $n$.

At first sight, there appears to be a conflict between the locality of
the emission of a baby universe, and hence the locality of the
non-conservation of matter energy, and the existence, in the gauge 
(\ref{4.1*}), of the long ranged dilaton field whose strength is
determined by the matter energy. To see that this conflict is only apparent, 
let us present a scenario consistent with both of the above properties.
We stress that the following consideration is only a scenario, as its 
confirmation or rejection would require the analysis of the final state of
the parent universe, which goes well beyond the scope of this paper.
Also, the discussion below is essentially classical, while the actual
analysis should necessarily be at the quantum level.

Let us again consider the collision of two narrow pulses of matter, and choose
the gauge (\ref{4.2*}). In this gauge the field $\rho$ and the matter energy
density are those shown in fig.3 and fig.6a. If the matter energy was 
conserved, the final state would be characterized by the configuration 
shown in fig.6b by dashed lines: the field $\rho$ between the pulses
would change from $-\rho_0$ to $+\rho_0$ where $\rho_0$ is determined 
by the total matter energy in the pulses (see eqs.(\ref{17c}) and
(\ref{13a}), (\ref{13d})). If the energy is not conserved at the moment
of the collision  (i.e., if the collision of the pulses induces splitting 
off of the baby universe), the 
height of the matter pulses and, correspondingly,
the height of the plateau of $\rho$ are smaller in the final state;
this configuration is shown in fig.6b by solid lines.

Clearly, the process shown in fig.6 may be perfectlly local in 
(1+1)-dimensional space-time. It shows  that non-conservation of 
matter energy does not require non-locality. However, this process {\it cannot
be transformed into the gauge} (\ref{4.1*}), as the field $\rho$  
 does not obey the field equation $\partial_{\alpha} \partial^{\alpha} \rho =0$
everywhere in space--time. To see what happens if the gauge (\ref{4.1*}) is
chosen for the {\it initial} state, let us perform the gauge transformation
that would transform the ``conventional'' configuraion (i.e., the
configuration of the conventional process with energy conservation)
into the gauge (\ref{4.1*}). In the case of infinitely narrow pulses
this gauge transformation is the inverse of \eq{17d}. Then the initial state
is one shown in fig.2 (with $\rho= {\rm const} \cdot \sigma^0$ everywhere),
while the final state is that shown in fig.7 by solid lines 
(only a small region of the universe is presented in fig.7; 
the final configuration
 of the conventional proccess with energy conservation
is again shown  by dashed lines for comparison).  The final dilaton 
field $\phi$ in this gauge is the same as that of   the conventional process;
in particular, its long range behavior  is not affected by 
energy non-conservation. On the other hand, the field $\rho$ in the final 
state is non-trivial and corresponds to longitudinal gravitational waves.
It is the presence of these longitudinal waves that ensures the validity of
the constraints after the collision, even  though the energy 
of matter is not conserved and the dilaton field does not change
asymptotically. Of course, the longitudinal $\rho$-wave may be gauged away,
but the corresponding gauge transformation would be non-trivial far away from
the collision region, and would also induce longitudinal gravitational
wave in the initial state.

In infinite space, the gauge (\ref{4.2*}) cannot be imposed, so we cannot
use the arguments based on fig.6. However, the final states like  
those shown in fig.7 are still possible in the gauge where $\rho=0$
initially. For these final states to appear, the field equations should be
violated only in a small region of space-time (where the two pulses collide),
and the entire process may occur locally. The ADM mass viewed from  infinite
distance is conserved, but this conservation is due to the
appearance of the longitudinal gravitational waves that compensate for
non-conservation of matter energy.

It remains to be understood what part, if any, of the discussion
of this paper may be relevant to (3+1)-dimensional theories. There
exist semiclassical arguments, based on the study of fluctuations 
about\cite{Shvedov} and analytical continuation of\cite{LRTcont}
the Euclidean wormhole solution of ref.\cite{GS1}, 
favouring the interpretation
of the wormhole as describing the process in which a baby universe
branches off and then ``flies away'' in (mini)superspace. This process may be
very similar to the one discussed in this paper  in (1+1)-dimensional context.
On the other hand, the possible non--conservation of bare energy was not
explicit in the semiclassical treatment of (3+1)-dimensional Euclidean 
wormholes.
An independent problem which can possibly be treated ``phenomenologically'',
as we did in this section, is whether non-conservation of bare energy
in (3+1) dimensions is consistent with locality, and, in particular, 
whether locality requires the generation of longitudinal gravitational waves.
We hope our study of (1+1)-dimensional toy model will be helpful to
understand  these problems.

The author is indebted to A.A. Tseytlin for very helpful correspondence
and to P.G. Tinyakov for useful discussions. This work is supported in 
part by the
Russian Foundation for Basic Research, project 96-02-17449a, by INTAS 
grant 93-1630-ext and CRDF grant 649.

\vskip7mm

{\bf \large Appendix}
\vskip3mm
Let us outline the calculation of the matrix elements (\ref{2.5*}) and
(\ref{2.6+}) in the leading order in ${\bf f}_n$, 
$\tilde{{\bf f}}_{\tilde{n}}$. First, we make the projection onto the
subspace $L_0 = \tilde{L}_0$ explicit by writing the coherent state 
(\ref{2.5+}) in the following form,
\begin{equation}
   \ket{\Psi_P} = \int_{-\pi/2}^{\pi/2}~ \frac{d\xi}{\pi}~
             \exp\left( \sum_n \frac{1}{n} {\bf f}_n(\xi) {\bf A}_{-n} +
               \sum_{\tilde{n}} \frac{1}{\tilde{n}} 
               \tilde{{\bf f}}_{\tilde{n}}(\xi) \tilde{{\bf A}}_{-\tilde{n}}
               \right)
               \ket{{\cal P}}
\label{A.1*}
\end{equation}
where
\[
        {\bf f}_n(\xi) = {\bf f}_n \e^{2in\xi}
\]
\begin{equation}
 \tilde{{\bf f}}_{\tilde{n}}(\xi) = \tilde{{\bf f}}_{\tilde{n}} 
  \e^{-2i\tilde{n}\xi}
\label{A.1+}
\end{equation}
The representation (\ref{A.1*}) coincides with \eq{2.5+} up to normalization.
The fact that the state (\ref{A.1*}) obeys the constraint 
$(L_0 - \tilde{L}_0)\ket{\Psi_P} = 0$ follows from the commutational 
relations of the DDF operators with $L_0$ and $\tilde{L}_0$,
\[
      [L_0,A_n^i] = -\frac{n}{2} A_n^i
\]
\[
      [\tilde{L}_0,A_n^i] = \frac{n}{2} A_n^i
\]
and similarly for $\tilde{A}^j_{\tilde{n}}$.

Consider now the norm of these states. One has
\[
   \braket{\Psi_{P'}}{\Psi_P} = \left[
         \int~\frac{d\xi_1 d\xi_2}{\pi^2} \exp \left( \sum
          {\bf f}_n {\bf f}_n^{*} \e^{2in(\xi_1 - \xi_2)} +\sum
          \tilde{{\bf f}}_{\tilde{n}} \tilde{{\bf f}}_{\tilde{n}}^{*} 
           \e^{-2i\tilde{n}(\xi_1 - \xi_2)} \right) \right]
          \braket{{\cal P}'}{{\cal P}}
\]
At large ${\bf f}_n$ and $\tilde{{\bf f}}_{\tilde{n}}$ this is a saddle point
integral. Taking into account \eq{2.5**} we find that the integrand does not
depend on $(\xi_1 + \xi_2)$, while the saddle point in $(\xi_1 - \xi_2)$
is at 
\begin{equation}
    \xi_1 - \xi_2 = 0
\label{A.2+}
\end{equation}
Hence we obtain the usual result
\[
     \braket{\Psi_{P'}}{\Psi_P} = \braket{{\cal P}'}{{\cal P}}
           \exp \left( \sum
          {\bf f}_n {\bf f}_n^{*}  +\sum
          \tilde{{\bf f}}_{\tilde{n}} \tilde{{\bf f}}_{\tilde{n}}^{*} 
            \right)
\]
up to a pre-exponential factor.

Let us turn to the matrix elements (\ref{2.5*}) involving matter fields
only and consider explicitly the left-moving sector. The DDF operators
can be written as follows,
\begin{eqnarray}
  A^i_n &=& \int_{-\pi/2}^{\pi/2}~ \frac{d \sigma_{+}}{\pi}
         \exp \left(2in \frac{x^{(+)}}{P^{(+)}} + 2in\sigma_{+}
              - \frac{2n}{P^{(+)}} \sum \frac{1}{k} \alpha_{k}^{(+)}
                \e^{-2ik\sigma_{+}} \right)  \nonumber \\
&& \nonumber \\
&& \times
           \left( \frac{1}{2} P^i + \sum \alpha^i_q \e^{-2iq\sigma_{+}}
           \right)
\label{A.2*}
\end{eqnarray}
Since $P^i$ and $P^{(+)}$ commute with $A^i_n$, one can set
\[
     P^i = {\cal P}^i,~~~~~~~ P^{(+)} = {\cal P}^{(+)}
\]
in the operator $O_M$ for calculating the matrix element
(\ref{2.5*}). Furthermore,  $\alpha_k^{(+)}$ commute with
$O_M$ and with all factors in \eq{A.2*}, so one can set them equal to zero
and write effectively
\[
      A^i_n = \exp \left( 2in \frac{x^{(+)}}{P^{(+)}} \right) \alpha^i_n
\]
We have to calculate the matrix element
\begin{equation}
\bra{\Psi_{P'}} \alpha^{i_1}_{-r_1} \dots \alpha^{i_s}_{-r_s} \cdot
   \alpha^{j_1}_{p_1} \dots \alpha^{j_t}_{p_t} \ket{\Psi_P}
\label{A.3*}
\end{equation}
with $r_1,\dots,r_s,p_1,\dots,p_t >0$, which is a building block of \eq{2.5*}.
Note that the operator ordering in \eq{A.3*} is in fact not essential at
large ${\bf f}_n$, as is usual in the classical limit. One finds for
this matrix element
\begin{eqnarray}
        \int~\frac{d\xi_1 d\xi_2}{\pi^2}&& \bra{{\cal P}'}
           \exp\left[ 2i(r_1 + \dots + r_s - p_1 - \dots - p_t)
            \frac{x^{(+)}}{{\cal P}^{(+)}}\right]\cdot ({\rm RM})
            \ket{{\cal P}} \nonumber \\
&& \nonumber \\
&& \times
           f^{*}_{r_1}(\xi_2)\dots f^{*}_{r_s}(\xi_2)\cdot
          f_{p_1}(\xi_1)\dots f_{p_t}(\xi_1) \cdot ({\rm RM})
\label{A.4+}
\end{eqnarray}
where we omitted the superscripts $i_1,\dots,j_t$; (RM) denotes the 
corresponding factors due to right-moving modes. All dependence on
$(\xi_1 + \xi_2)$ in this integral comes from the exponents in
$f^{*}_r (\xi_2)$ and $f_p (\xi_1)$, see \eq{A.1+}, and similar exponents 
for right-moving components. This makes the matrix element in \eq{A.4+}
equal to $\braket{{\cal P}'}{{\cal P}}$. The integral over $(\xi_1 - \xi_2)$
is still of saddle point structure with the saddle point (\ref{A.2+}).
Hence the expression (\ref{A.4+}) simplifies and becomes equal to
\[
     \braket{\Psi_{P'}}{\Psi_P} \cdot \int~ \frac{d\xi}{\pi}~
      f^{*}_{r_1}(\xi)\dots f^{*}_{r_s}(\xi)\cdot
          f_{p_1}(\xi)\dots f_{p_t}(\xi) \cdot ({\rm RM})
\]
We conclude that in the leading order in
${\bf f}_n$, $\tilde{{\bf f}}_{\tilde{n}}$,
the calculation of the matrix elements of the matter operators is
reduced to the substitution
\[
      \alpha_n \to \e^{-2in\xi}f_n~,~~~~n>0
\]
\[
      \alpha_{-n} \to \e^{2in\xi}f_n^{*}~,~~~~n>0
\]
with subsequent integration over $\xi$. This proves the relation between 
the matrix elements (\ref{2.5*}) and their
classical counterparts (\ref{2.7*}).

Let us turn to the matrix elements (\ref{2.6+}). Since $P^{(+)}$ and
$\alpha_k^{(+)}$ commute with $O_M$ and with the DDF operators, the operator
$\partial_{+} X_{L}^{(+)}$ reduces to
\begin{equation}
    \partial_{+} X_{L}^{(+)} = \frac{1}{2} {\cal P}^{(+)}
\label{A.5+}
\end{equation}
when sandwitched as in \eq{2.6+}. To find the matrix elements involving 
$\partial_{+} X_{L}^{(-)}$ one notices that in the leadig order in ${\bf f}_n$,
$\tilde{{\bf f}}_{\tilde{n}}$
\begin{equation}
      \bra{\Psi_{P'}} L_m O_M \ket{\Psi_P} = 0
\label{A.5*}
\end{equation}
because the commutator of $L_m$ and $O_M$ does not contain
 ${\bf f}_n$, $\tilde{{\bf f}}_{\tilde{n}}$ (recall that $\ket{\Psi_P}$ and
$\ket{\Psi_{P'}}$ are physical states). Equation (\ref{A.5+}) and just 
established  relation between the matrix elements (\ref{2.5*}) and classical
correlators (\ref{2.7*}) immediately imply the desired relation between
the matrix elements (\ref{2.6+}) and their classical versions (\ref{2.7+}).
This relation can of course be obtained by an explicit calculation.

\begin{figure}[p]
\unitlength=1.00mm
\special{em:linewidth 0.4pt}
\linethickness{0.4pt}
\begin{picture}(150.00,148.00)
\put(80.00,106.00){\vector(0,1){34.00}}
\bezier{312}(90.00,109.00)(94.00,148.00)(98.00,109.00)
\bezier{28}(98.00,109.00)(98.00,106.00)(102.00,106.00)
\bezier{28}(90.00,109.00)(89.00,106.00)(85.00,106.00)
\bezier{312}(62.00,109.00)(66.00,148.00)(70.00,109.00)
\bezier{28}(70.00,109.00)(70.00,106.00)(74.00,106.00)
\bezier{28}(62.00,109.00)(61.00,106.00)(57.00,106.00)
\put(85.00,137.00){\makebox(0,0)[cc]{$\varepsilon_M$}}
\put(150.00,101.00){\makebox(0,0)[cc]{$\sigma^1$}}
\put(80.00,27.00){\vector(0,1){70.00}}
\put(69.00,72.00){\line(1,0){22.00}}
\put(150.00,79.00){\makebox(0,0)[cc]{$\sigma^1$}}
\put(83.00,95.00){\makebox(0,0)[cc]{$\phi$}}
\put(10.00,85.00){\vector(1,0){140.00}}
\put(10.00,106.00){\vector(1,0){140.00}}
\put(99.00,68.00){\line(6,-5){43.00}}
\put(140.00,32.33){\line(0,0){0.00}}
\put(61.00,68.00){\line(-6,-5){43.00}}
\bezier{36}(61.00,68.00)(66.00,72.00)(69.00,72.00)
\bezier{48}(89.00,72.00)(95.00,72.00)(99.00,68.00)
\end{picture}
\centerline{Fig.~1.}
\end{figure}

\begin{figure}[p]
\unitlength=1.00mm
\special{em:linewidth 0.4pt}
\linethickness{0.4pt}
\begin{picture}(150.00,148.00)
\put(80.00,106.00){\vector(0,1){34.00}}
\bezier{312}(90.00,109.00)(94.00,148.00)(98.00,109.00)
\bezier{28}(98.00,109.00)(98.00,106.00)(102.00,106.00)
\bezier{28}(90.00,109.00)(89.00,106.00)(85.00,106.00)
\bezier{312}(62.00,109.00)(66.00,148.00)(70.00,109.00)
\bezier{28}(70.00,109.00)(70.00,106.00)(74.00,106.00)
\bezier{28}(62.00,109.00)(61.00,106.00)(57.00,106.00)
\put(85.00,137.00){\makebox(0,0)[cc]{$\varepsilon_M$}}
\put(150.00,101.00){\makebox(0,0)[cc]{$\sigma^1$}}
\put(80.00,27.00){\vector(0,1){70.00}}
\put(69.00,72.00){\line(1,0){22.00}}
\put(150.00,79.00){\makebox(0,0)[cc]{$\sigma^1$}}
\put(83.00,95.00){\makebox(0,0)[cc]{$\phi$}}
\put(10.00,85.00){\vector(1,0){140.00}}
\put(10.00,106.00){\vector(1,0){140.00}}
\put(140.00,32.33){\line(0,0){0.00}}
\bezier{36}(61.00,68.00)(66.00,72.00)(69.00,72.00)
\bezier{48}(89.00,72.00)(95.00,72.00)(99.00,68.00)
\put(99.00,68.00){\line(1,-1){16.00}}
\put(61.00,68.00){\line(-1,-1){15.00}}
\bezier{124}(115.00,52.00)(125.00,40.00)(140.00,40.00)
\bezier{108}(46.00,53.00)(34.00,39.00)(25.00,40.00)
\put(24.00,89.00){\makebox(0,0)[cc]{$\pi/2$}}
\put(136.00,89.00){\makebox(0,0)[cc]{$\pi/2$}}
\put(136.00,110.00){\makebox(0,0)[cc]{$\pi/2$}}
\put(24.00,110.00){\makebox(0,0)[cc]{$\pi/2$}}
\put(24.00,106.00){\line(0,1){2.00}}
\put(24.00,85.00){\line(0,1){2.00}}
\put(136.00,85.00){\line(0,1){2.00}}
\put(136.00,106.00){\line(0,1){2.00}}
\end{picture}
\centerline{Fig.~2.}
\end{figure}

\begin{figure}[p]
\unitlength=1.00mm
\special{em:linewidth 0.4pt}
\linethickness{0.4pt}
\begin{picture}(150.00,148.00)
\put(80.00,106.00){\vector(0,1){34.00}}
\bezier{312}(109.00,109.00)(113.00,148.00)(117.00,109.00)
\bezier{28}(117.00,109.00)(117.00,106.00)(121.00,106.00)
\bezier{28}(109.00,109.00)(108.00,106.00)(104.00,106.00)
\bezier{312}(47.00,109.00)(51.00,148.00)(55.00,109.00)
\bezier{28}(55.00,109.00)(55.00,106.00)(59.00,106.00)
\bezier{28}(47.00,109.00)(46.00,106.00)(42.00,106.00)
\put(85.00,137.00){\makebox(0,0)[cc]{$\varepsilon_M$}}
\put(150.00,101.00){\makebox(0,0)[cc]{$\sigma^1$}}
\put(80.00,27.00){\vector(0,1){70.00}}
\put(150.00,79.00){\makebox(0,0)[cc]{$\sigma^1$}}
\put(83.00,95.00){\makebox(0,0)[cc]{$\rho$}}
\put(10.00,85.00){\vector(1,0){140.00}}
\put(10.00,106.00){\vector(1,0){140.00}}
\put(140.00,32.33){\line(0,0){0.00}}
\put(24.00,89.00){\makebox(0,0)[cc]{$\pi/2$}}
\put(136.00,89.00){\makebox(0,0)[cc]{$\pi/2$}}
\put(136.00,110.00){\makebox(0,0)[cc]{$\pi/2$}}
\put(24.00,110.00){\makebox(0,0)[cc]{$\pi/2$}}
\put(24.00,106.00){\line(0,1){2.00}}
\put(24.00,85.00){\line(0,1){2.00}}
\put(136.00,85.00){\line(0,1){2.00}}
\put(136.00,106.00){\line(0,1){2.00}}
\put(10.00,82.00){\line(1,0){25.00}}
\bezier{84}(35.00,82.00)(45.00,82.00)(50.00,72.00)
\bezier{84}(50.00,72.00)(56.00,62.00)(65.00,62.00)
\put(62.00,62.00){\line(1,0){36.00}}
\bezier{88}(98.00,62.00)(108.00,62.00)(114.00,72.00)
\bezier{80}(114.00,72.00)(119.00,82.00)(128.00,82.00)
\put(126.00,82.00){\line(1,0){24.00}}
\end{picture}
\centerline{Fig.~3.}
\end{figure}

\begin{figure}[p]
\unitlength=0.8mm
\special{em:linewidth 0.8pt}
\linethickness{0.8pt}
\begin{picture}(124.00,121.00)(0,20)
\bezier{156}(104.00,97.00)(116.00,81.00)(104.00,66.00)
\bezier{16}(101.00,97.00)(103.00,98.00)(104.00,97.00)
\bezier{16}(101.00,66.00)(103.00,65.00)(104.00,66.00)
\bezier{156}(39.00,97.00)(51.00,81.00)(39.00,66.00)
\bezier{152}(36.00,97.00)(25.00,82.00)(36.00,66.00)
\bezier{16}(36.00,97.00)(38.00,98.00)(39.00,97.00)
\bezier{16}(36.00,66.00)(38.00,65.00)(39.00,66.00)
\put(57.00,91.00){\circle*{1.50}}
\put(92.00,91.00){\circle*{1.50}}
\put(113.00,70.00){\vector(1,0){11.00}}
\put(11.00,70.00){\vector(1,0){11.00}}
\put(19.00,78.00){\makebox(0,0)[cb]{P}}
\put(117.00,77.00){\makebox(0,0)[cb]{P}}
\put(103.00,65.50){\line(-1,0){65.00}}
\put(38.00,97.50){\line(1,0){65.00}}
\bezier{20}(99.00,93.00)(98.00,91.00)(97.00,88.00)
\bezier{28}(100.00,69.00)(98.00,72.00)(97.00,75.00)
\bezier{24}(96.00,84.00)(95.00,81.00)(96.00,78.00)
\bezier{28}(92.00,91.00)(95.00,93.00)(92.00,95.00)
\bezier{28}(92.00,95.00)(89.00,97.00)(92.00,99.00)
\bezier{28}(92.00,99.00)(95.00,101.00)(92.00,103.00)
\bezier{24}(92.00,103.00)(90.00,105.00)(92.00,107.00)
\bezier{24}(92.00,113.00)(90.00,115.00)(92.00,117.00)
\bezier{28}(92.00,117.00)(95.00,119.00)(92.00,121.00)
\put(57.00,91.00){\circle*{1.50}}
\bezier{28}(57.00,91.00)(60.00,93.00)(57.00,95.00)
\bezier{28}(57.00,95.00)(54.00,97.00)(57.00,99.00)
\bezier{28}(57.00,99.00)(60.00,101.00)(57.00,103.00)
\bezier{24}(57.00,103.00)(55.00,105.00)(57.00,107.00)
\bezier{24}(57.00,113.00)(55.00,115.00)(57.00,117.00)
\bezier{28}(57.00,117.00)(60.00,119.00)(57.00,121.00)
\put(52.00,105.00){\vector(0,1){13.00}}
\put(98.00,118.00){\vector(0,-1){12.00}}
\put(45.00,112.00){\makebox(0,0)[cc]{Q}}
\put(106.00,112.00){\makebox(0,0)[cc]{Q}}
\bezier{32}(92.00,107.00)(95.00,110.00)(92.00,113.00)
\bezier{32}(57.00,107.00)(60.00,110.00)(57.00,113.00)
\centerline{Fig.~4.}
\end{picture}
\end{figure}

\begin{figure}[p]
\unitlength=0.8mm
\special{em:linewidth 0.8pt}
\linethickness{0.8pt}
\begin{picture}(124.00,121.00)(0,20)
\bezier{156}(104.00,97.00)(116.00,81.00)(104.00,66.00)
\bezier{16}(101.00,97.00)(103.00,98.00)(104.00,97.00)
\bezier{16}(101.00,66.00)(103.00,65.00)(104.00,66.00)
\bezier{156}(39.00,97.00)(51.00,81.00)(39.00,66.00)
\bezier{152}(36.00,97.00)(25.00,82.00)(36.00,66.00)
\bezier{16}(36.00,97.00)(38.00,98.00)(39.00,97.00)
\bezier{16}(36.00,66.00)(38.00,65.00)(39.00,66.00)
\put(57.00,91.00){\circle*{1.50}}
\put(92.00,91.00){\circle*{1.50}}
\put(113.00,70.00){\vector(1,0){11.00}}
\put(11.00,70.00){\vector(1,0){11.00}}
\put(19.00,78.00){\makebox(0,0)[cb]{P}}
\put(117.00,77.00){\makebox(0,0)[cb]{P'}}
\put(103.00,65.50){\line(-1,0){65.00}}
\put(38.00,97.50){\line(1,0){65.00}}
\bezier{20}(99.00,93.00)(98.00,91.00)(97.00,88.00)
\bezier{28}(100.00,69.00)(98.00,72.00)(97.00,75.00)
\bezier{24}(96.00,84.00)(95.00,81.00)(96.00,78.00)
\bezier{28}(92.00,91.00)(95.00,93.00)(92.00,95.00)
\bezier{28}(92.00,95.00)(89.00,97.00)(92.00,99.00)
\bezier{28}(92.00,99.00)(95.00,101.00)(92.00,103.00)
\bezier{24}(92.00,103.00)(90.00,105.00)(92.00,107.00)
\bezier{24}(92.00,113.00)(90.00,115.00)(92.00,117.00)
\bezier{28}(92.00,117.00)(95.00,119.00)(92.00,121.00)
\put(57.00,91.00){\circle*{1.50}}
\bezier{28}(57.00,91.00)(60.00,93.00)(57.00,95.00)
\bezier{28}(57.00,95.00)(54.00,97.00)(57.00,99.00)
\bezier{28}(57.00,99.00)(60.00,101.00)(57.00,103.00)
\bezier{24}(57.00,103.00)(55.00,105.00)(57.00,107.00)
\bezier{24}(57.00,113.00)(55.00,115.00)(57.00,117.00)
\bezier{28}(57.00,117.00)(60.00,119.00)(57.00,121.00)
\put(52.00,105.00){\vector(0,1){13.00}}
\put(98.00,118.00){\vector(0,-1){12.00}}
\put(45.00,112.00){\makebox(0,0)[cc]{Q}}
\put(106.00,112.00){\makebox(0,0)[cc]{K}}
\bezier{32}(92.00,107.00)(95.00,110.00)(92.00,113.00)
\bezier{32}(57.00,107.00)(60.00,110.00)(57.00,113.00)
\end{picture}

\centerline{Fig.~5}
\end{figure}

\begin{figure}[p]
\begin{center}
\psfig{file=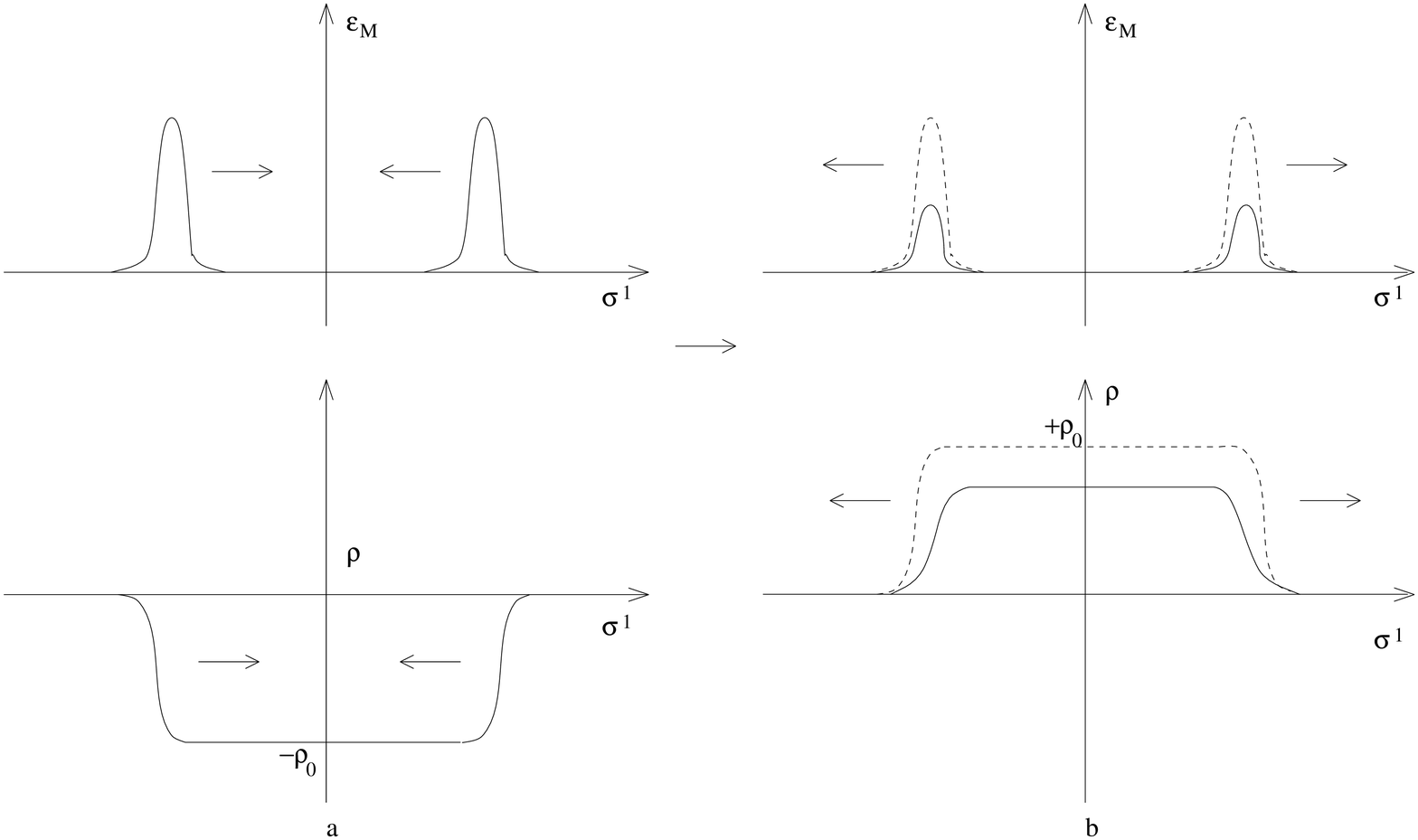,width=6in}
Fig.~6.
\end{center}
\end{figure}

\begin{figure}[p]
\begin{center}
  \psfig{file=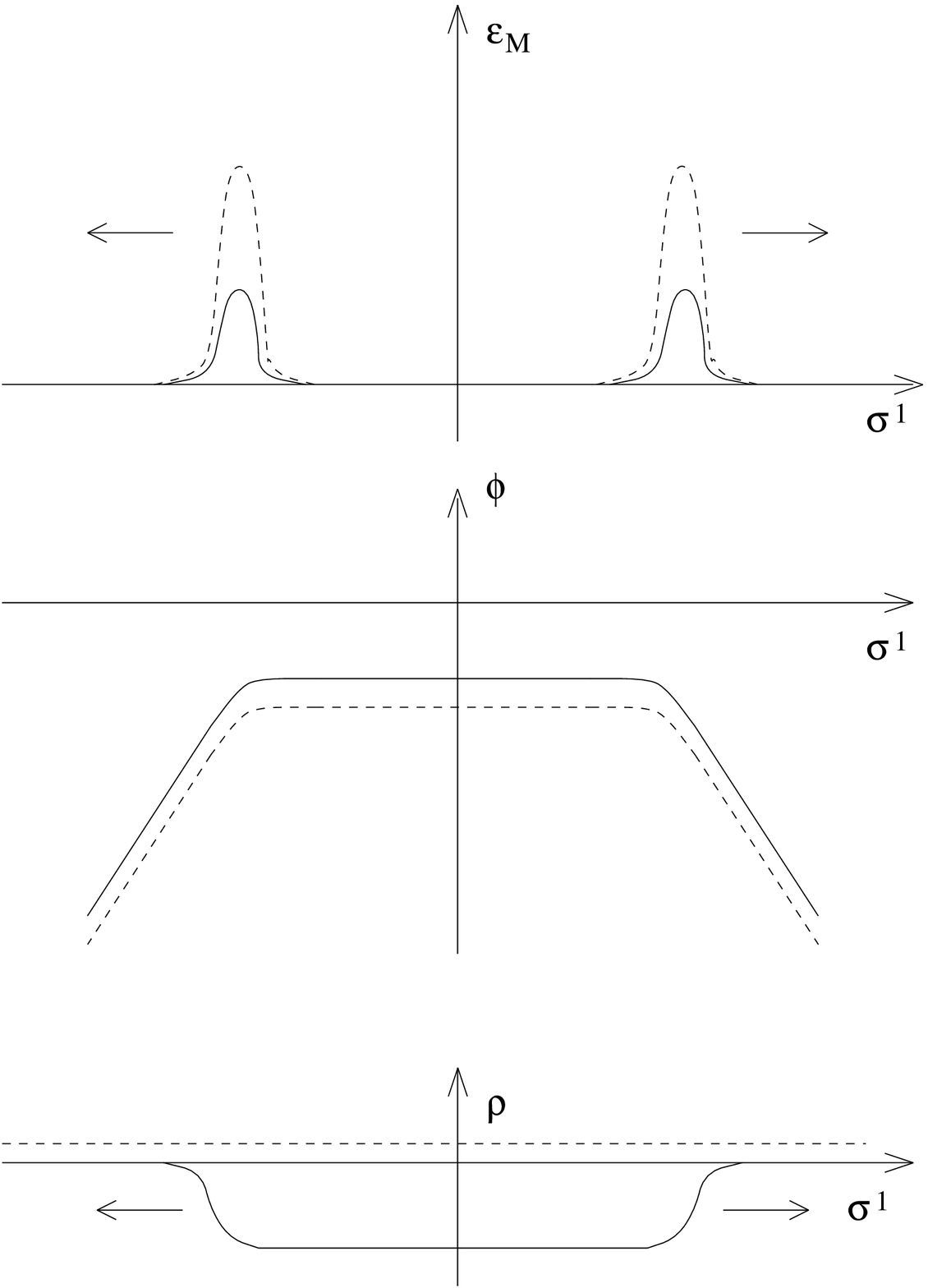,width=5in}
  Fig.~7.
\end{center}
\end{figure}


\begin{thebibliography}{99}
\bibitem{Ginsparg} P. Ginsparg and G. Moore, {\it Lectures on 2-D
gravity and 2-D string theory}, TASI Summer School, Boulder, 1992.
\bibitem{StromingerReview} A. Strominger, {\it Les Houches lectures
          on black holes}, Les Houches Summer School, Les Houches, 1994.
\bibitem{Hawking1}S. Hawking, {\em Phys.Lett.} {\bf 195B} (1987) 337;\\
                 in: Quantum Gravity. Proc. 4th Int. Seminar, Moscow, 1987,
		 eds.M.A.Markov, V.A.Berezin and V.P.Frolov (World
                 Scientific, 1988).
\bibitem{LRT}G.V. Lavrelashvili, V.A. Rubakov and P.G.Tinyakov, {\em JETP
                  Lett.} {\bf 46} (1987) 167;\\
                 in: Quantum Gravity. Proc. 4th Int. Seminar, Moscow, 1987,
		 eds.M.A.Markov, V.A.Berezin and V.P.Frolov (World
                 Scientific, 1988).
\bibitem{GS1} S.B.Giddings and A.Strominger, {\em Nucl.Phys.}
                  {\bf B306} (1988) 890.
\bibitem{Coleman1} S. Coleman, {\em Nucl.Phys.} {\bf B307} (1988) 890.
\bibitem{GS2} S.B. Giddings and A. Strominger, {\em Nucl.Phys.}
                  {\bf B307} (1988) 854.
\bibitem{Banks}T. Banks, {\em Physicalia Magazine} {\bf 12 suppl} (1990) 19.
\bibitem{Hawking2}S.W. Hawking, {\em Nucl.Phys.} {\bf B363} (1991) 117.
\bibitem{LyHa}A. Lyons and S.W. Hawking, {\em Phys.Rev.} {\bf D44} (1991)
                   3802.
\bibitem{VR} V.A. Rubakov,
          {\em Nucl. Phys.} {\bf B453} (1995) 395.
\bibitem{Nirov} Kh.S. Nirov and V.A. Rubakov, {\em Mod. Phys. Lett.}
        {\bf A10} (1995) 3059.
\bibitem{BanksPS} T. Banks, L. Susskind and M.E. Peskin,
          {\em Nucl. Phys.} {\bf B244} (1984) 125;\\
                  M. Srednicki,
                  {\em Nucl. Phys.} {\bf B410} (1993) 143.
\bibitem{EllisEnergy} J. Ellis, N.E. Mavromatos and D.V. Nanopoulos,
    preprint CERN-TH-6755-92, {\em Int. J. Mod. Phys.} {\bf A11} (1996) 1489;\\
{\it A noncritical string approach to black holes, time
and quantum dynamics}, Erice Lectures 1993, CERN-TH-7195-94 (1994)
\bibitem{Unruh} W.G. Unruh and R.M. Wald, {\em Phys. Rev.} {\bf D52} 
            (1995) 2176
\bibitem{VRFilippov} V.A. Rubakov,
 {\it Quantum coherence, energy conservation and baby universes: lessons 
from (1+1) dimensions}, in Problems of Theoretical Physics,
Dedicated to 60th Birthday of A.T. Filippov (JINR, Dubna, 1996)
\bibitem{CGHS} C.G. Callan, S.B. Giddings, J.A. Harvey and A. Strominger,
             {\em Phys. Rev.} {\bf D45} (1992) R1005
\bibitem{BilalKogan} A. Bilal and I.I. Kogan, {\em Phys. Rev.} {\bf D47}
              (1993) 5408
\bibitem{Bilal} A. Bilal, {\em Int. J. Mod. Phys.} {\bf A9} (1994) 475
\bibitem{Tseytlin1} A.A. Tseytlin, {\em Nucl. Phys.} {\bf B390} (1993) 153
\bibitem{Tseytlin2}A.A. Tseytlin, {\em Phys. Rev.} {\bf D47} (1993) 3421
\bibitem{Verlinde1} K. Schoutens, H. Verlinde and E. Verlinde,
         {\em Phys. Rev.} {\bf D48} (1993) 2670
\bibitem{Verlinde2} E. Verlinde and H. Verlinde, 
              {\em Nucl. Phys.} {\bf B406} (1993) 43
\bibitem{Kazama1} S. Hirano, Y. Kazama and Y. Satoh,
{\em Phys. Rev.} {\bf D48} (1993) 1687
\bibitem{Kazama2} Y. Kazama and Y. Satoh,
{\em Phys. Rev.} {\bf D50} (1994) 2368; 3889
\bibitem{Russo} J. Russo and A. Tseytlin,
{\em Nucl. Phys.} {\bf B382} (1992) 259
\bibitem{TseytlinPrivate} A.A. Tseytlin, private communication.
\bibitem{DDF}E. Del Guidice, P. Di Veccia and S. Fubini,
                {\em Ann. of Phys.} {\bf 70} (1972) 378.
\bibitem{Shvedov} V.A. Rubakov and O.Yu. Shvedov, {\em Phys. Lett.}
            {\bf 383} (1996) 258; \\
         in: Proc. Int. Seminar ``Quarks-96'',
       Yaroslavl, 1996, to appear, gr-qc/9608065.
\bibitem{LRTcont} G.V. Lavrelashvili, V.A. Rubakov and P.G. Tinyakov,
{\em Mod. Phys. Lett.} {\bf A3} (1988) 1231;\\
V.A. Rubakov and
P.G.  Tinyakov, {\em Phys.Lett.} {\bf B214} (1988) 334.


%\bibitem{KSB}I. Klebanov, L. Susskind and T. Banks, {\em Nucl.Phys.}
%                  {\bf B317} (1989) 665.

%\bibitem{GSW}M.B.Green, J.H.Schwarz and E.Witten,
%	       Superstring theory (Cambridge U. Press, Cambridge, 1987).
%\bibitem{Dualamplitudes} K.Bardak\c{c}i and H.Ruegg, {\em Phys. Rev.}
%                  {\bf 181} (1969) 1884.
\end{thebibliography}
\end{document}